\documentclass[runningheads]{llncs}
\usepackage[misc]{ifsym}
\usepackage{enumitem}
\usepackage{graphicx}
\usepackage{algorithm}
\usepackage{algorithmicx}
\usepackage{algpseudocode}
\usepackage{multirow}
\usepackage{float}
\usepackage{booktabs}
\usepackage{xcolor}
\usepackage{url}
\usepackage{color}
\usepackage{array}
\usepackage{tikz}
\usepackage{hyperref}
\usepackage{colortbl}
\newcommand{\tabincell}[2]{\begin{tabular}{@{}#1@{}}#2\end{tabular}}
\definecolor{mycyan}{rgb}{0.88,0.90,0.94}
\begin{document}
\title{DuCN: Dual-children Network for Medical Diagnosis and Similar Case Recommendation towards COVID-19%\\A Diagnosis and Similar Case Recommendation Model for COVID-19% \thanks{Supported by organization x.}
}
% \author{Authors}
\author{Chengtao Peng$^{1}$$^{(\scriptsize\textrm{\Letter})}$, Yunfei Long$^2$, Dandan Tu$^2$, Senhua Zhu$^2$, Bin Li$^1$}
\authorrunning{CPeng et al.}
\institute{Affiliations}
\institute{1 Department of Electronic Engineering and Information Science, University of Science and Technology of China, Hefei, China \\
2 EI Innovation Lab, Huawei, Shanghai, China\\
Email: pct@mail.ustc.edu.cn
% 3 Department of Computer Science and Engineering, University of Notre Dame, Notre Dame, Indiana, USA
}
\maketitle              % typeset the header of the contribution
\begin{abstract}
Early detection of the coronavirus disease 2019 (COVID-19) helps to treat patients timely and increase the cure rate, thus further suppressing the spread of the disease. In this study, we propose a novel deep learning based detection and similar case recommendation network to help control the epidemic. Our proposed network contains two stages: the first one is a lung region segmentation step and is used to exclude irrelevant factors, and the second is a detection and recommendation stage. Under this framework, in the second stage, we develop a dual-children network (DuCN) based on a pre-trained ResNet-18 to simultaneously realize the disease diagnosis and similar case recommendation. Besides, we employ triplet loss and intrapulmonary distance maps to assist the detection, which helps incorporate tiny differences between two images and is conducive to improving the diagnostic accuracy. For each confirmed COVID-19 case, we give similar cases  to provide radiologists with diagnosis and treatment references. We conduct experiments on a large publicly available dataset (CC-CCII) and compare the proposed model with state-of-the-art COVID-19 detection methods. The results show that our proposed model achieves a promising clinical performance.

\end{abstract}
\section{Introduction}
Up to the present, the coronavirus disease 2019 (COVID-19) has caused massive infections and deaths over the world, and is still mutating rapidly.
% and unfortunately, the related viruses (SARS-CoV-2~\cite{world2020laboratory}) are still mutating.
To contain the pandemic, it is urgently demanded to find an efficient way for COVID-19 detection and symptom analysis. % to prevent the further spread of the disease while diminishing the mortality rate. 
In the clinical practice, the accurate detection of COVID-19 mainly relies on manual diagnosis using chest X-ray computed tomography (CT) images since they can show early infected lesions. Woefully, manual detection calls for highly-professional experiences and intensive labours, which makes it difficult for radiologists to screen for the disease. Therefore, developing an automated computer-aided COVID-19 detection system could effectively alleviate radiologists' burden and assist the disease detection.
% the detection of COVID-19 mainly relies on reverse transcription-polymerase
% chain reaction (RT-PCR) and chest imaging (e.g., chest X-ray computed tomography (CT)). However, RT-PCR usually shows deficient sensitivity to test specimens with low viral load present or laboratory errors~\cite{bai2020performance}. In consequence, some countries used chest imaging (especially CT, which can show early lesions in the lung), as a first-line investigation and patient management tool. If a radiologist is full experienced, relying on CT images detection may obtain high accuracy. Woefully, due to the fact that COVID-19 is a new disease and the number of cases is huge, coupled with the lack of relevant experience and heavy workload of radiologists, it is difficult to accurately screen for the disease.
% Thus, developing an automated computer-aided COVID-19 detection system is very helpful for radiologists. 

In recent years, deep learning (DL) based methods has made great achievements in lung disease analysis~\cite{guan2020thorax,liang2019dense,wang2019thorax,liu2020kiseg}. Also, many excellent works were proposed for COVID-19 detection~\cite{bai2020artificial,hemdan2020covidx,javaheri2020covidctnet,wang2020covid} since 2019. The common DL-based COVID-19 detection methods usually employed a convolutional neural network (CNN) to extract image features and yielded predictions. For example, Jin et al.~\cite{jin2020development} designed a deep CNN detection system to diagnose COVID-19. % and achieved AUC of 92.99\% and 93.25\% on two different public available datasets: CC-CCII~\cite{zhang2020clinically} and MosMedData~\cite{morozov2020mosmeddata}, respectively.
Gao et al.~\cite{gao2021dual} developed a dual-branch combination network, which combined the related lesion attention maps to assist the detection. Zhang et al.~\cite{zhang2020clinically} proposed a diagnosis framework, which detected COVID-19 from other common pneumonias (CP) and normal healthy cases based on segmented lesion regions and achieved satisfactory results. %the area under receiver operating characteristic curve (AUC) score of 0.9797. 
Minaee et al.~\cite{minaee2020deep} used four pretrained models (ResNet18~\cite{he2016deep}, ResNet50~\cite{he2016deep}, SqueezeNet~\cite{iandola2016squeezenet} and DenseNet-121~\cite{huang2017densely}) and deep transfer learning technique to do the detection. Ter-Sarkisov~\cite{ter2020lightweight} introduced a lightweight Mask R-CNN~\cite{he2017mask} model to reduce the number of network parameters in COVID-19 detection. Hu et al.~\cite{hu2020weakly} developed a weakly supervised multi-scale learning framework, which assimilates different scales of lesion information for COVID-19 detection.

\begin{figure}[t]
\includegraphics[width=\textwidth]{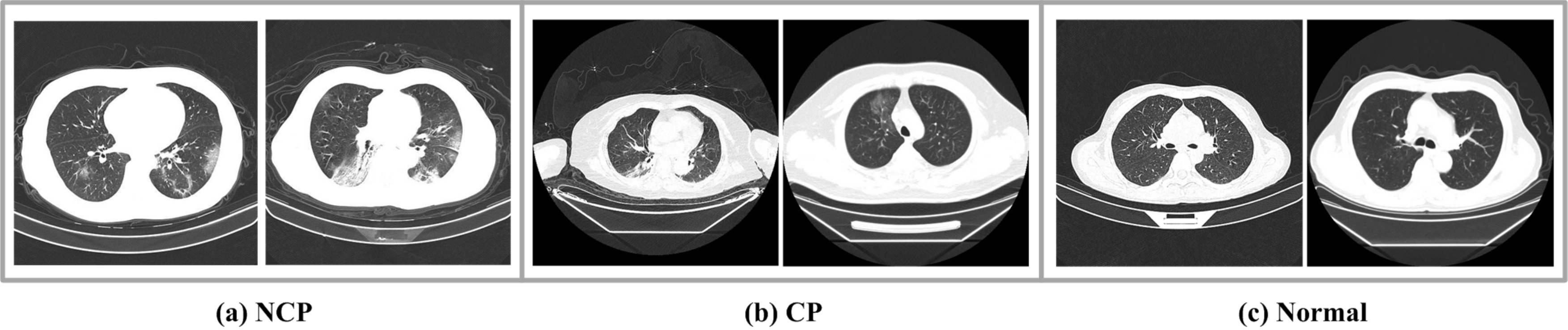}
\caption{Illustration of images with different diseases. NCP, CP, Normal mean norvel coronavirus pneumonias, common pneumonias and normal healthy cases, respectively.} \label{different_images}
\end{figure}

The above-mentioned methods were demonstrated to be effective and contributed a lot to combat COVID-19. However, the methods were proposed based on 2D images, which may omit important inter-slice information. To better extract representative features of COVID-19 lesions, several works~\cite{li2020using,li2020efficient,ouyang2020dual,wang2020weakly,zheng2020deep} proposed to diagnose the disease using 3D images. Among these approaches, Ouyang et al.~\cite{ouyang2020dual} proposed to focus on the infection regions inside lungs, and then conducted the detection; while other works directly extracted features from entire images and make predictions. 3D methods usually outperforms 2D models since they incorporated more spatial features (e.g., inter-slice information) from 3D images. However, known 3D models were usually trained using the common cross-entropy loss, which may be strenuous to extract tiny differences between the novel coronary pneumonia (NCP) and CP% (e.g., see Fig.~\ref{different_images} (a-b), there is very little difference between the CT images of the lungs infected with NCP and CP)
. Moreover, none of the existing works provides similar cases for confirmed COVID-19 case, while similar cases may provide radiologists with significant treatment references.

In view of this, in this paper, we propose a dual-children network (DuCN) to simultaneously detect COVID-19 and provide similar cases for the confirmed case. In the proposed model, lung regions are segmented at first to exclude irrelevant regions (see Fig~\ref{different_images} (a), the COVID-19 infection regions are mainly inside the lungs), hence eliminating the interference of irrelevant factors. Then, the segmented lung images are used for detection and recommendation. Meanwhile, the corresponding original full CT images and Euclidean distance maps in the lung regions are also integrated for providing abundant information. For COVID-19 detection process, we apply the triplet loss~\cite{schroff2015facenet} to extract slight differences between NCP and CP. At last, once a case is confirmed with COVID-19, we bestow radiologists with similar cases, providing the diagnostic evidence and treatment references. Verified  on a large clinical publicly available dataset (CC-CCII~\cite{zhang2020clinically}) and compared with state-of-the-art methods, our new method yields promising results for COVID-19 detection and similar case recommendation.

Compared with previous COVID-19 detection works, our work makes the following contributions: (1) We develop a DuCN, which could give similar cases at the same time of diagnosis. To our best knowledge, this is the first work to provide COVID-19 similar cases; (2) We propose to use a triplet loss to supervise network to extract tiny differences between different type of images in COVID-19 detection; (3) We propose to use intrapulmonary Euclidean distance maps to assist incorporate more spatial information of infected lesions.

Our code is made publicly available at GitHub (Anonymous Information).

% \begin{itemize}
% \item We develop a dual-children network, which could give similar cases at the same time of diagnosis. To our best knowledge, this is the first work to provide COVID-19 similar cases.
% \item We propose to transfer the problem of COVID-19 detection into a recommendation problem, which is conducive to extracting more representative information and improving diagnosis accuracy.
% \item We propose to use intrapulmonary Euclidean distance maps to assist the COVID-19 detection/recommendation, which may be helpful to incorporate more spatial information of infected lesions.
% \end{itemize}

\section{Method}
\subsection{Proposed Model}

Generally, the infection of the novel coronavirus is mainly occurs in the lung area, and has little effect on areas outside lungs. Hence, we divide the detection and recommendation process into two stages: lung segmentation and detection/recommendation.

\begin{figure}[t]
\centering                                        
\includegraphics[width=\textwidth]{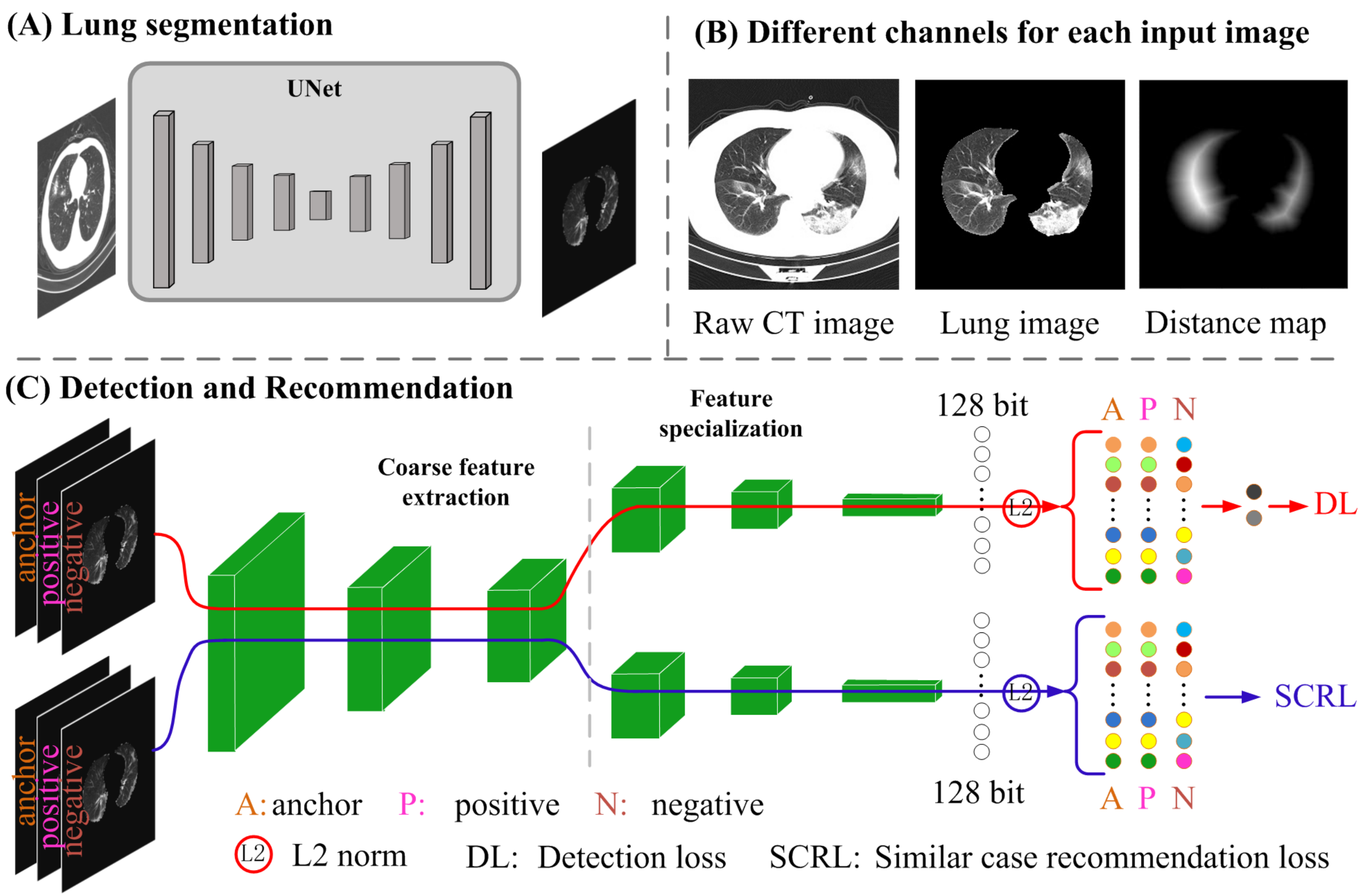}
\caption{Illustrating our proposed DuCN for medical diagnosis and similar case recommendation towards COVID-19.} \label{network}
\end{figure}

Fig.~\ref{network} gives an overview of our proposed model. In the framework, firstly the original CT images are resized from $512 \times 512$ to $224 \times 224$ and input into a lung segmentation network, which is constructed by the common U-Net~\cite{ronneberger2015u}, and used to produce lung masks. The segmented lung masks are then combined with the corresponding original CT images to produce lung images. Besides, Euclidean distance maps in the lung regions are computed based on the segmented masks as in the previous work~\cite{liang2019dense}. Finally, the dual-children network takes the original CT images, lung images and intrapulmonary Euclidean distance maps as inputs, and yields the probability of being infected with COVID-19. If a case is confirmed with COVID-19, the image-level similar cases are further provided. Below we present more details of our proposed model.

Someone may be conscious that our network has two more inputs (original CT images and intrapulmonary Euclidean distance maps) than previous works (e.g.,~\cite{wang2020weakly}), which advocated using lung images to detect COVID-19. In general, if patients are infected with the novel coronary pneumonia, especially for critical patients, their liver will also be damaged~\cite{fan2020clinical}. Thus, the liver morphology in original CT images may promote the detection of the COVID-19. %In addition, for critical patients with COVID-19, the infected area of the lung may be connected to the pleura, making it difficult to completely segment the lung area. 
In this situation, using the full original CT images may be help make up for the information loss caused by probable inaccurate lung segmentation. For the extra input of intrapulmonary Euclidean distance maps, they are conducive to extracting more spatial information of infected lesions and improving the diagnosis/recommendation accuracy.

\subsection{Dual-children Network}
\label{DuCN}

The existing detection networks usually directly output the possibility of infection, without providing relevant diagnostic evidence or similar cases. However, in clinical practice, the reference of similar cases is of great significance to the treatment of diseases, especially for a new disease such as COVID-19. With this in mind, in this study, we develop a dual-children network (DuCN) to simultaneously detect COVID-19 and provide similar cases for confirmed patients.

Fig.~\ref{network} (C) shows the proposed DuCN. As shown, the DuCN has two paths: the red path is designed for disease detection and the blue one is developed for similar case recommendation. For each path, we employ a pretrained ResNet18~\cite{he2016deep} as backbone except that the output digits (1000) is replaced by 128. The two paths share network parameters in coarse feature extraction phase (the first three levels in ResNet18), and work independently in feature specialization phase (see Fig.~\ref{network} (C)). The reason for this design is that the shallow features of a network mainly contain high-frequency information such as gray scale, which are universal for both detection and recommendation. While the deep-level features of a network mainly contain low-frequency information (e.g., semantic details), which is desired differently for various tasks. Therefore, when the DuCN goes deep, the paths for different tasks work separately and do not share parameters.

Please note that in the training and testing phases, the DuCN works differently. Below we elucidate the details for two different tasks respectively.

% \noindent
% \textbf{COVID-19 detection}
\subsubsection{COVID-19 detection}
For COVID-19 detection, the previous works tended to use the common CrossEntropy loss to guide feature extraction. However, the CrossEntropy loss may be not adept at picking up tiny differences between different types (some NCP and CP cases have very similar lesions). Triple loss~\cite{schroff2015facenet}, which was exploited for face recognition, could explore delicate changes between different cases due to the more comprehensive comparison between different type of cases. In this work, except for the CrossEntropy loss, to improve the detection sensitivity of the network, we employ the triplet loss to guide the delicate feature extraction. The triplet loss is defined as:
\begin{equation}
L_{triplet}=max\{d(f(a),f(p))-d(f(a),f(n))+margin,0\}
\end{equation}
where $d$ represents the Euclidean distance; $f$ is a feature extractor; $a$ means an anchor sample, which is randomly selected from COVID-19 cases; $p$ is an image (positive sample) of the same type as $a$, and is also randomly selected from COVID-19 cases, but from a different patient from $a$; $n$ is a different type of negative sample from $a$, randomly selected from non-COVID-19 cases (including CP and normal cases in this study); $margin$ is the distance between positive and negative samples after training, we set it is to 1.2 according to our numerical experiments.

The network for COVID-19 detection is presented in Fig.~\ref{network} (C) (the red path). In the training phase, the input for this task is triplet images ($a$, $p$ and $n$), each of which contains three channels: lung image, full original CT image, and intrapulmonary Euclidean distance map. Each image in triplet images is processed by a ResNet18 (the red path in Fig.~\ref{network} (C)) in turn and outputs 128-bits one-dimensional features (3*128-bits in total). The extracted features are further normalized by an $L_2$ regularization and used for computing triplet loss. Then the features are transferred in 2-bits representations by a linear layer and used to compute CrossEntropy loss. After training, in the testing phase, the image to be detected (contains 3-channels) is input into the network and generate two scores that represent the probability of different diseases.

\vspace{-12pt}
\subsubsection{Similar case recommendation}
In clinical practice, doctors usually consult related similar confirmed cases to diagnose and treat existing cases, especially for some emerging or intractable diseases (e.g., COVID-19). Thus, devising an automated system to provide similar confirmed cases would be helpful for treating diseases. In this study, we develop a network to provide similar confirmed COVID-19 cases to help control the epidemic. The network is shown in Fig.~\ref{network} (C) (the blue path). In the training phase, the input is triplet images as in the detection path, the differences for the recommendation test are that $p$ is randomly selected from the same patient as $a$, and $n$ is randomly selected from a different patient but with the same disease (COVID-19) as $a$. This path is 
supervised by triplet loss only. After training, all images in the dataset are transferred into a 128-bit representations and saved as 
an index database. When an image is confirmed as COVID-19, it would be transferred into a 128-bit representations though the trained model and compared with samples in the library one by one using the Euclidean distance. The one with the smallest distance is the most similar case.

\vspace{-5pt}
\subsection{Loss Functions}
Our proposed full network model has two sub-networks: U-Net and DuCN. The two sub-networks are trained separately, the segmentation network (U-Net) is trained using the common Dice loss, and DuCN is trained via the loss: %$L_{total}=0.3*L_{D}+0.7*L_{SCR}$.
\begin{equation}
L_{total}=0.3*L_{D}+0.7*L_{SCR}
\label{total_loss}
\end{equation}
where $L_{SCR}=L_{triplet}$ is the similar case recommendation loss, and $L_{D}$ is the COVID-19 detection loss, which is defined as:
% \begin{equation}
% L_{CrossEntropy}=-\sum_{i=1}^{N}(p_k*log(q_k))
% \end{equation}
\begin{equation}
L_{D}=0.4*L_{triplet}+0.6*L_{CrossEntropy}
\label{detection_loss}
\end{equation}
where $L_{CrossEntropy}$ represents the common CrossEntropy loss. The scale factors in loss functions are set based on our experience and numerical experiments.%(Eq. (\ref{total_loss} - \ref{detection_loss})) 

\vspace{-5pt}
\section{Experiments and Results}

\subsection{Dataset and Experiments}
For lung segmentation, we use a public dataset~\cite{ma_jun_2020_3757476}, which was collected for COVID-19 lung and lesion segmentation, to train the network. The dataset %contains 4,240 CT slices and the corresponding lung masks, which 
is split into 70\% and 30\% for training and testing, respectively. 
For COVID-19 detection and similar case recommendation, we employ a large dataset (CC-CCII~\cite{zhang2020clinically}) to evaluate the proposed DuCN. %The datasets consisted of a total 361,221 slices from 2,246 patients including 752 NCP patients, 797 CP patients and 697 normal control patients. 
We use data augmentation (flip left and right, and rotate $\pm 2^\circ$) to expand the number of NCP images to a level equivalent to the number of non-NCP images. We split the dataset in patient-level (prevent CT images of the same patient from appearing simultaneously in training and testing sets) into 70\% and 30\% for training and testing, respectively. The quantitative evaluation for this study includes: dice score (for segmentation); sensitivity, specificity, precision, accuracy, F-1 score and AUC (for detection); and subjective judgments (for segmentation and similar case recommendation).%, we only use subjective judgments to evaluate the performance.

In our experiments, the two networks (U-Net and DuCN) are both implemented by PyTorch (V1.2.0) and trained separately on four NVIDIA K80 GPUs (12GB) with a batch size of 256. We adopt the Adam optimizer to train the networks with an initial learning rate of $1  \times 10^{-3}$, divided by 2 every 20 epochs.% The two networks are both implemented by PyTorch and trained on four NVIDIA K80 GPUs (12GB) with a batch size of 256.

\vspace{-5pt}
\subsection{Results}
\subsubsection{Lung Segmentation}

\begin{figure}[ht]
\includegraphics[width=\textwidth]{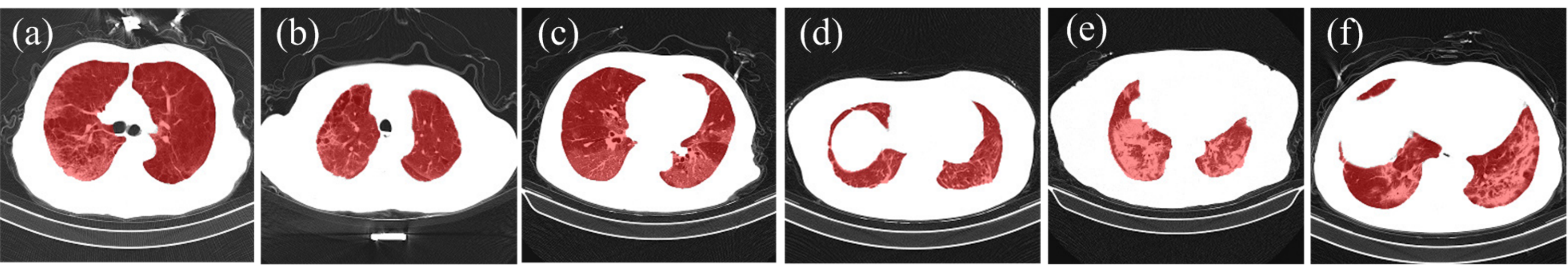}
\caption{Some representative visual segmentation results for lung regions.} \label{segmentation_results}
\end{figure}

\begin{table}[t]
\renewcommand\tabcolsep{5pt}
\centering
\caption{Quantitative results and comparison of different models. All the values are computed based on the entire testing dataset and are reported on a percentage scale.}
\label{tab1}
\begin{tabular}{ccccccc}
\toprule[1pt]
Methods & Sensitivity & Specificity & Precision & F1-score & Accuracy & AUC\\
\toprule[1pt]
\rowcolor{mycyan}
\cite{zhang2020clinically} & 94.93 &91.13 &N/A &N/A &92.49 &\textbf{97.97}\\

\tabincell{c}{CovidNet-S~\cite{he2021automated}} & 91.72 & N/A &88.78 &90.23 &88.55 &N/A\\
\rowcolor{mycyan}
\tabincell{c}{CovidNet-L~\cite{he2021automated}} & 88.08 & N/A &90.48 &89.26 &88.69 &N/A\\

\cite{wu2021covid} & N/A & N/A & N/A & N/A & 86.60 & 96.80\\

\rowcolor{mycyan}
\tabincell{c}{\cite{jin2020development}}& N/A & N/A & N/A & N/A & N/A & 92.99\\

% \rowcolor{mycyan}
% \tabincell{c}{\cite{nguyen2021deep}} & N/A &N/A & N/A & N/A & N/A &98.80\\

% \toprule[0.5pt]

\tabincell{c}{DuCN-LMR} & 96.32 & 91.28 &95.46 & 95.89& 95.21 &93.96\\
\rowcolor{mycyan}
\tabincell{c}{DuCN-DMR} & 99.67 & \textbf{94.43} &98.13 & 98.89& 97.86 &96.98\\
\tabincell{c}{DuCN-RIR} & 98.63 & 94.25 &97.82 & 98.22& 97.90 &96.88\\
% \tabincell{c}{DuCN-TL} & 93.08 & 90.49 &99.64 & 99.51& 99.24 &99.03\\
\rowcolor{mycyan}
\tabincell{c}{DuCN-UP} & 99.68 & 94.40 &98.07 & 98.86& 98.01 &96.92\\
\tabincell{c}{Proposed} & \textbf{99.85} & 94.41 & \textbf{98.54}& \textbf{98.90} & \textbf{98.28} &97.13\\

\toprule[1pt]
\end{tabular}
\vspace{-15pt}
\end{table}

Lung segmentation plays a critical role in the following step for COVID-19 detection and similar case recommendation. In the first stage, our segmentation network produces results with a dice score of 0.9669 on the entire testing dataset. However, since the dataset used for training segmentation network is different from the dataset in the second stage, a high dice score does not mean that the segmentation model will perform well in the second stage. Therefore, we apply the trained segmentation network in CC-CCII and evaluate the performance.

Fig.~\ref{segmentation_results} shows some representative visual segmentation results. Subjectively, the trained segmentation network performs very well, even for critical patients (e.g., see Figs.~\ref{segmentation_results} (e-f)). The accurate segmentation of lung regions provides accurate and comprehensive lesion information for the subsequent step, and further promotes the accurate diagnosis of the disease.

\vspace{-15pt}
\subsubsection{COVID-19 Detection}
In this work, we devise a DuCN to do COVID-19 detection and similar case recommendation. In this section, to verify the performance of the detection, we compare the results with five state-of-the-arts~\cite{he2021automated,jin2020development,wu2021covid,zhang2020clinically}, among which~\cite{he2021automated} developed two models (CovidNet-S and CovidNet-L). The comparison methods were all conducted on the same publicly dataset: CC-CCII. Table~\ref{tab1} shows the values of various metrics for different methods, to ensure the fairness, we refer to the values reported in the original papers (N/A means that the original works did not use this indicator). From the comparison, it is clear that our new model produces promising results for COVID-19 detection.

\vspace{-15pt}
\subsubsection{Similar Case Recommendation}

\begin{figure}[t]
\includegraphics[width=\textwidth]{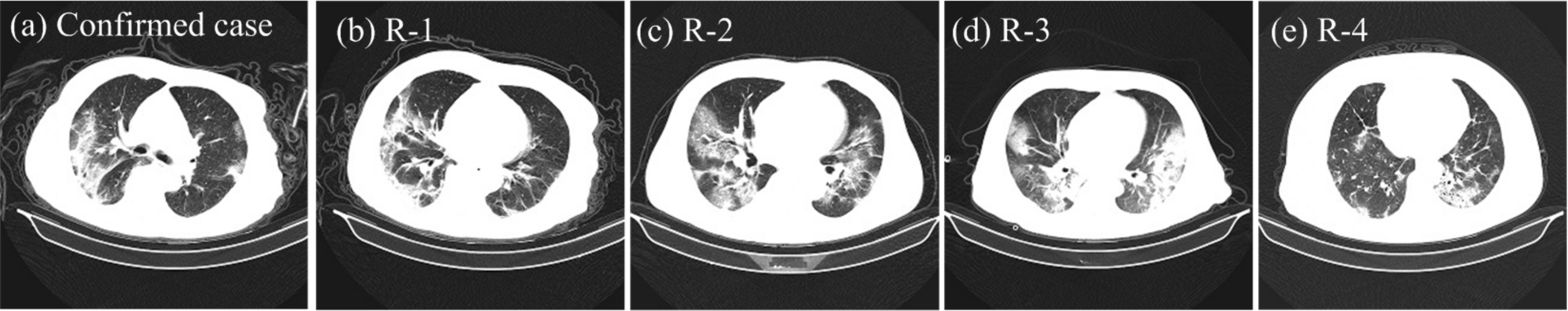}
\caption{Illustration of top-4 similar cases for confirmed COVID-19 case. R-N means the recommended case ranked N.} \label{recommendation_results}
\vspace{-5pt}
\end{figure}

For confirmed cases, it is of great significance for treatment if doctors are provided relevant case references. Thus, our proposed DuCN gives similar cases at the same time of diagnosis. Since it is hard to evaluate such function quantitatively, we only assess it subjectively. Fig.~\ref{recommendation_results} presents a confirmed COVID-19 patient and its top-5 relevant cases. The top 1 is the patient itself, and for other three recommendations, the cases are related to the confirmed patient% (assessed by three radiologists from Anonymous hospital)
, implying that our similar case recommendation system is effective.

\vspace{-5pt}
\subsection{Ablation Study}
\subsubsection{Effectiveness of Raw Images, Lung masks and Distance Maps} For each slice of image that input into DuCN, it has 3-channels (a raw image, a lung image and a intrapulmonary distance map). To show the effectiveness of the 3-channels, we remove one of them from input. To ensure the input still contains 3 channels, if raw image or distance map removed (indicated as DuCN-RIR/DuCN-DMR), we replaced it with the corresponding lung image; if lung mask removed (indicated as DuCN-LMR), lung image and distance map are both replaced by the corresponding raw image. The results are illustrated in Table~\ref{tab1}, from the results, all the three revisions has negative effect, suggesting that raw images, lung images and intrapulmonary Euclidean distance maps are all useful.

% \subsubsection{Effectiveness of Triplet Loss}
% In the COVID-19 detection task, we use a triplet loss to assist DuCN to extract tiny differences between different type of images. To  validate  the effectiveness  of it, we remove it and only use the crossentropy loss to train the network. The results of the revised model is presented in Table~\ref{tab1} (DuCN-TL), as seen, the performance is degraded. Thus, the triplet loss does help to improve our DuCN.
\vspace{-12pt}
\subsubsection{Does Pretrained Network Help?}
In this study, we use a pretrained ResNet-18 to construct DuCN. To test if the pretrained model help, we replace it in the full network with an untrained ResNet-18. The results is listed in Table~\ref{tab1} (DuCN-UP), one may observe that the pretrained network achieves a higher performance than the untrained one. Thus, it is reasonable to choose the pretrained ResNet18 to construct our DuCN.

\vspace{-5pt}
\section{Discussion and Conclusions}
In this paper, we proposed a new DuCN to provide similar cases for confirmed COVID-19 patients at the same time of diagnosis. To exclude the interference of irrelevant factors outside lungs, we used a segmentation network to segment lung regions. Besides, the original CT images were incorporated to extract lesion related features (e.g., liver information). Further, intrepulmonary distance maps and triplet loss were introduced to help extract tiny differences. Validated on a large public dataset, our proposed network exhibits a promising performance in clinical application. Our proposed DuCN can be generically used in other disease screening applications. Still, our method has some shortcomings. For instance, for similar case recommendation, we selected positive samples from the same patient as anchor images for network training, which was not religious. Strictly, patients with similar infected lesions and symptoms should be regarded as similar cases. In future work, we will make effort to improve this.

\bibliographystyle{splncs04}
%\bibliography{References} 

\bibliography{reference}
\end{document}